\PassOptionsToPackage{protrusion, expansion}{microtype}

\documentclass[conference, nofonttune]{IEEEtran}

\usepackage[maxbibnames=99]{biblatex}

\usepackage{comment}
\usepackage{amsmath, amsthm, amssymb, mathtools, bm} 
    
\usepackage{graphicx, xcolor}               
    \definecolor{B}    {HTML}{2b66d3}
    \definecolor{B2}   {HTML}{003399}
    \definecolor{R}    {HTML}{c9171e}
    \definecolor{R2}   {HTML}{d7003a}
    \definecolor{INK}  {HTML}{595857}
    \definecolor{Y}    {HTML}{f1c40f}
    \definecolor{G}    {HTML}{009a00}
    \definecolor{GRAY} {HTML}{808080}
    \definecolor{MAUVE}{HTML}{9400D1}
\usepackage{url, hyperref}                  
	\hypersetup{
	   pdfborder={0 0 0}}
\usepackage{adjustbox}                      
\usepackage{booktabs, multirow}             
\usepackage{caption, subcaption}            
\usepackage{enumitem}                       
\usepackage{lipsum}                         
\usepackage{setspace}                       
\usepackage[boxed]{algorithm}       		
    \floatstyle{plaintop}
\usepackage{algpseudocode}                  
    \algrenewcommand{\alglinenumber}[1]{{\scriptsize\bfseries\ttfamily\color{R}#1}}
    \algrenewcommand\algorithmiccomment[1]{\hfill{\color{gray}\fontsize{6.5}{7}\selectfont\sffamily$\triangleright$\ #1}}
    
\makeatletter
\newcommand\fs@ruled@nomiddle{\def\@fs@cfont{\bfseries}\let\@fs@capt\floatc@ruled
    \def\@fs@pre{}%
    \def\@fs@mid{\vskip 1pt\hrule height.8pt depth0pt \kern2pt}%
    \def\@fs@post{\vskip 5pt\hrule height.8pt depth0pt \relax}%
  \let\@fs@iftopcapt\iftrue}
\renewcommand\fst@algorithm{\fs@ruled@nomiddle}
\makeatother

\usepackage{xpatch}
\makeatletter
    \xpatchcmd{\algorithmic}{\ALG@tlm\z@}{\ALG@tlm\z@\leftmargin 10pt}{}{}
\makeatother

\usepackage{footnote}                       
\usepackage{listings}						
\usepackage{colortbl}
\usepackage{tikz}
\usetikzlibrary{positioning}
\newcommand*\Circled[1]{
	\tikz[baseline=(char.base)]{\node[
        shape=circle, draw=none,  thick, 
        fill=gray!40,inner sep=0.6pt] (char) 
    {\textcolor{black}{\sffamily#1}}; 
}}

\usepackage{array}
\usepackage{rotating}

\newcommand{\YES}{$\bullet$}
\newcommand{\TabTitle}[1]{%
	\multicolumn{1}{@{}c@{}}{\rotatebox{90}{#1}}}
\newcommand{\TabTitleTL}[2]{%
	\multicolumn{1}{@{}c@{}}{\rotatebox{90}{%
	\begin{minipage}{6em}#1\\[-.8ex]#2\end{minipage}
	}}}

\newcommand{\SyncBlk}{sync block}
\newcommand{\SyncGrd}{sync grid}
\newcommand{\SyncDev}{sync device}

\newcommand{\NullCell}{\multicolumn{1}{c}{}}

\newcommand{\TITLE}{\sffamily\bfseries}
\newcommand{\EvalTableTitle}[3]{%
    \multicolumn{#1}{#2}{%
        \sffamily\bfseries #3}%
    }
\newcommand{\Unit}[1]{\textcolor{black!66}{\normalfont\sffamily\scshape #1}}

\newcommand{\DatumSize}[2]{{\scriptsize #1} \Unit{#2}}
\newcommand{\gpuRTX}{%
    \raisebox{1.pt}{%
        \bfseries\scshape\sffamily%
        \setlength{\fboxsep}{1.5pt}%
        \colorbox{gray!30}{tu}%
    }
}
\newcommand{\gpuVolta}{%
    \raisebox{1.pt}{%
        \sffamily\scshape\bfseries%
        \setlength{\fboxsep}{1.5pt}%
        \colorbox{black}{\color{white}v}%
    }
}

\newcommand{%
    \tableHeadBold}[1]{\multicolumn{1}{c}{\sffamily\bfseries\begin{tabular}{@{}c@{}}#1\end{tabular}}%
}

\newcommand{%
    \notAvailable}{{\setlength{\fboxsep}{1pt}\fbox{$\times$}}%
}

\newcommand{\BOLD}{\fontfamily{ugq}\selectfont}

\newcommand{\func}[1]{{\ttfamily#1}}
\newcommand{\var}[1]{{\ttfamily#1}}

\hypersetup{pdfauthor={J. Tian}, pdftitle={Revisiting Huffman Coding}}

\newcommand{\SEC}{{\S}}

\newcommand{\TAB}{{Table}}

\newcommand{\TABFONTFAMILY}{\ttfamily}

\usepackage[scale=.97]{tgheros} 
\usepackage[scale=.95]{newtxtt}

\IEEEoverridecommandlockouts

\begin{document}

\title{Revisiting Huffman Coding: Toward Extreme Performance on Modern GPU Architectures}

\newcommand{\BetterMark}{$^\star$}

\author{Jiannan Tian\BetterMark, 
Cody Rivera\IEEEauthorrefmark{2},
Sheng Di\IEEEauthorrefmark{3},
Jieyang Chen\IEEEauthorrefmark{4},
Xin Liang\IEEEauthorrefmark{4},
Dingwen Tao{\BetterMark}\thanks{Corresponding author: Dingwen Tao (\url{dingwen.tao@wsu.edu}), School of EECS, Washington State University, Pullman, WA 99164, USA.},
and Franck Cappello\IEEEauthorrefmark{3}\IEEEauthorrefmark{5}\\
\BetterMark%
School of Electrical Engineering and Computer Science, Washington State University, WA, USA\\
\IEEEauthorrefmark{2}Department of Computer Science, The University of Alabama, AL, USA\\
\IEEEauthorrefmark{3}Mathematics and Computer Science Division, Argonne National Laboratory, IL, USA\\
\IEEEauthorrefmark{4}Oak Ridge National Laboratory, TN, USA\\
\IEEEauthorrefmark{5}University of Illinois at Urbana-Champaign, IL, USA
}

\maketitle

\thispagestyle{plain}
\pagestyle{plain}

\begin{abstract}
Today's high-performance computing (HPC) applications are producing vast volumes of data, which are challenging to store
and transfer efficiently during the execution, such that data compression is becoming a critical technique to mitigate
the storage burden and data movement cost.
Huffman coding is arguably the most efficient Entropy coding algorithm in information theory, such that it could be
found as a fundamental step in many modern compression algorithms such as DEFLATE. 
On the other hand, today's HPC applications are more and more relying on the accelerators such as GPU on supercomputers,
while Huffman encoding suffers from low throughput on GPUs, resulting in a significant bottleneck in the entire data
processing. 
In this paper, we propose and implement an efficient Huffman encoding approach based on modern GPU architectures, which
addresses two key challenges: (1) how to parallelize the entire Huffman encoding algorithm, including codebook
construction, and (2) how to fully utilize the high memory-bandwidth feature of modern GPU architectures. The detailed
contribution is four-fold. (1) We develop an efficient parallel codebook construction on GPUs that scales effectively
with the number of input symbols. 
(2) We propose a novel reduction based encoding scheme that can efficiently merge the codewords on GPUs. (3) We optimize
the overall GPU performance by leveraging the state-of-the-art CUDA APIs such as Cooperative Groups. 
(4) We evaluate our Huffman encoder thoroughly using six real-world application datasets on two advanced GPUs and
compare with our implemented multi-threaded Huffman encoder.
Experiments show that our solution can improve the encoding throughput by up to 5.0$\times$ and 6.8$\times$ on NVIDIA
RTX 5000 and V100, respectively, over the  state-of-the-art GPU Huffman encoder, and by up to 3.3$\times$ over the
multi-thread encoder on two 28-core Xeon Platinum 8280 CPUs. 
\end{abstract}



\section{Introduction}
\label{sec:intro}

With the ever-increasing scale of HPC applications, vast volumes of data are produced during simulation, resulting in a
bottleneck for both storage and data movement due to limited capacity and I/O bandwidth. For example, Hardware/Hybrid
Accelerated Cosmology Code (HACC) \cite{hacc} (twice finalist nominations for ACM Gordon Bell Prize) produces 20
petabytes of data to store in one simulation of 3.5 trillion of particles with 300 timesteps, whereas leadership-class
supercomputers such as Summit \cite{summit} have limited storage capacities (around 50$\sim$200 PB) to be shared by
hundreds of users. On the other hand, network and interconnect technologies in HPC systems advance much more slowly than
computing power, causing intra-/inter-node communication cost and I/O bottlenecks to become a more serious issue in fast
stream processing \cite{use-case-Franck}. Compressing the raw simulation data at runtime and decompressing them before
post-analysis can significantly reduce communication and I/O overheads and hence improving working efficiency. 

Huffman coding is a widely-used variable-length encoding method that has been around for over 60 years
\cite{Huffman-original}. It is arguably the most cost-effective Entropy encoding algorithm according to information
theory, though some other coding methods such as arithmetic coding and range coding offer slightly better compression
ratios in a few specific cases. As such, Huffman coding algorithm serves as the critical step in many general-purpose
lossless compression software or libraries such as GZIP \cite{gzip}, Zstd \cite{zstd}, and Blosc \cite{blosc}. It is
also an integral part of many lossy compressors for image and video, such as JPEG \cite{jpeg}. Moreover, Huffman coding
is also extensively used in many error-bounded lossy compressors (such as SZ \cite{sz16,sz17} and MGARD
\cite{ainsworth2017mgard}), which have been very effective in compressing big scientific datasets with high data
fidelity, as verified by many existing studies \cite{liang2019improving, liang2019significantly, zhao2020significantly}.

In this paper, we focus on parallelizing the entire Huffman encoding algorithm on GPUs rather than the decoding stage.
On the one hand, the Huffman encoding algorithm plays a critical role in more and more HPC applications
\cite{Lal_Lucas_Juurlink_2017,choukse2020buddy} whose runtime performances are heavily relying on the GPU accelerators
of supercomputers. On the other hand, compared with decompression performance, compression performance (or encoding
efficiency) is particularly important to HPC applications, since large amounts of data need to be compressed on the fly
and poor compression performance may substantially counteract performance improvement resulting from the reduced data
size, causing inferior I/O performance \cite{liang2018error}. By contrast, decompression generally happens only during
the post-analysis, which has nothing to do with runtime performance of the simulation. 

However, there are no efficient Huffman encoding algorithms designed for the GPU, leaving a significant gap that needs
to be filled to meet modern HPC applications' requirements. Although many multi-thread Huffman encoding algorithms
already exist (e.g., Zstd extended its capability to run on multiple CPU cores \cite{zstd-multicore}), their design is
limited to coarse-grained parallelism, which is unsuitable for today's modern GPU architectures (featuring massive
single-instruction-multiple-thread (SIMT) mechanisms and high memory-bandwidth features). A few GPU-based Huffman
encoders, such as Rahmani et al.'s encoder \cite{Rahmani_Topal_Akinlar_2014}, adopt a rather simple parallel prefix sum
algorithm to calculate the location of each encoded symbol, which cannot fully utilize the GPU memory bandwidth due to
masses of movements of fragmented and variable-length data cells. Moreover, it is worth noting that traditionally,
Huffman coding is used in cases where there are 8 bits per symbol (i.e., 256 symbols in total), which is far less than
enough for many emerging HPC use cases. Error-bounded lossy compression, for example, often requires more than 8 bits
per codeword (e.g., 16 bits are required if $65536$ symbols are used in the codebook), because of potentially large
amount of integer numbers produced after the error-bounded quantization step \cite{sz17}. However, constructing a large
Huffman codebook sequentially may incur a significant performance bottleneck to the overall Huffman encoding on GPUs,
which was not addressed by any prior work. 

To address the significant gap, we present an efficient Huffman encoder on GPUs, which is compatible with many emerging
HPC scenarios. The basic idea is leveraging a battery of techniques to optimize performance on modern GPU architectures,
based on an in-depth analysis of Huffman encoding stage. Specifically, we optimize the parallel Huffman codebook
construction for GPUs and significantly reduce the overhead of constructing the codebook that involves a large number of
symbols. Moreover, we propose a novel reduction-based encoding scheme, which can significantly improve the memory
bandwidth utilization by iteratively merging codeword groups. To the best of our knowledge, our proposed and implemented
Huffman encoder is the first work that achieves hundreds of GB/s encoding performance on V100 GPU. The detailed
contributions are listed as follows. 
\begin{itemize}[noitemsep, topsep=2pt, leftmargin=1.3em]
\item We carefully explore how parallelization techniques can be applied to the entire Huffman encoding algorithm
    including histogramming, codebook construction, and encoding, and optimize each stage using state-of-the-art CUDA
    APIs such as Cooperative Groups. 
\item We develop an efficient parallel codebook construction on GPUs, especially for scenarios requiring a codebook with
    a large number of symbols, opening new possibilities for non-traditional use cases of Huffman coding.
\item We propose a novel reduction-based encoding scheme that iteratively merges the encoded symbols, significantly
    improving GPU memory bandwidth utilization. 
\item We evaluate our Huffman encoder on six real-world datasets using two state-of-the-art GPUs and compare it with
    other state-of-the-art Huffman encoders on both CPUs and GPUs. Experiments show that our solution can improve the
    encoding throughput by up to 6.8$\times$ on V100 and 3.3$\times$ on CPUs. 
\end{itemize}

In \SEC\ref{sec:bg}, we present the background for Huffman coding and parallel algorithms for its codebook construction.
In \SEC\ref{sec:ps}, we discuss the limitation of current Huffman encoding on GPUs.
In \SEC\ref{sec:design}, we present our proposed parallel Huffman codebook construction and reduction-based encoding scheme on GPUs. 
In \SEC\ref{sec:eval}, we show the experimental evaluation results. 
In \SEC\ref{sec:related} and \SEC\ref{sec:conclusion}, we discuss the related work and conclude our work.  

\section{Background}
\label{sec:bg}

\subsection{Huffman Coding and Its Emerging Applications}
\label{subs:bg-lossless}
Huffman coding is a fundamental data compression algorithm proposed by David Huffman in 1952 \cite{Huffman-original}. In
essence, it assigns codes to characters such that the length of the code depends on the relative frequency of the
corresponding character (a.k.a., input symbol). Huffman codes are variable-length and prefix-free. Here prefix-free
means no code is a prefix of any other. Any prefix-free binary code can be visualized as a binary tree (called the
Huffman tree) with the encoded characters stored at the leaves.

In recent years, data reduction attracts more and more attention in the HPC field, and Huffman coding becomes an integral part of many data reduction techniques such as error-bounded lossy compression \cite{sz16,zfp,ainsworth2017mgard}. 
For multiple HPC use cases, Huffman coding usually needs to be customized with a large number of symbols, instead of
using the classic Huffman coding with only 256 symbols/characters in the codebook. For example, SZ requires a customized
Huffman coding with $65536$ quantization bins in default, such that a large majority of integer codes generated by its
quantization could be covered by the encoding scheme. Such a customized Huffman coding is particularly critical when the
data is difficult to be predicted accurately, which is very common in scientific datasets.  

Another important scenario is n-gram compression \cite{liang2012segmenting}. For example, some languages have morphology
in the structure of words or morphemes, and it is important to utilize this syllable-based morphology for developing an
efficient text compression approach for these languages. Nguyen et~al. proposed a method~\cite{nguyen2016n} to partition
words into its syllables and then to produce their bit representations for compression. The number of bits in syllables
(symbols) depends on the number of entries in the dictionary file.  As another example, segmenting, encoding, and
decoding DNA sequences based on n-gram statistical language model is a critical research topic in bioinformatics to
handle the vast volumes of DNA sequencing data. Specifically, in this work \cite{liang2012segmenting}, researchers find
the length of most DNA words/symbols (e.g., 12$\sim$15 bits) and build an n-gram biology language model by analyzing the
genomes of multiple model species. Then, they design an approach to segment the DNA sequences and encode them
accordingly. 

In all the above cases, Huffman coding may require generating a codebook with a large number of symbols which is usually
far smaller than that of the input codewords. However, since such a large codebook is generated serially, codebook
construction can become a significant bottleneck, especially on small to medium-sized datasets. Thus it is vital to
develop an approach to build a Huffman codebook efficiently. 

\subsection{PRAM Model}
PRAM is a classic model to describe parallel algorithms where multiple processors are attached to a single memory
entity. It essentially assumes that \Circled{1} a set of processors of uniform type exist, and \Circled{2} all the
processors share a common memory unit with their accesses equal (via a memory access unit). This model is made
independent from specific hardware by introducing some ideal assumptions. The conventional taxonomy of read/write (R/W)
conflicts emphasize on the concurrency (denoted by C) and exclusiveness (denoted by E). Thus, there are four different
constraints that have been enforced on the PRAM model: EREW, ERCW, CREW, CRCW. In this paper, we focus on the CREW PRAM
model used by our parallel codebook construction algorithm and implement it on the GPU. 

\subsection{Parallel Huffman Codebook Construction}
\label{subs:bg-huffman-gen}

The serial Huffman codebook construction algorithm with the complexity of $\mathcal{O}(n\log n)$ constructs a
na\"{i}ve binary tree---a data structure not well-suited for the GPU memory. Specifically, the na\"{i}ve Huffman tree
has an inefficient GPU memory access pattern, which would incur a significant performance overhead on codebook
construction. This is confirmed by our experiment: constructing a Huffman codebook with 8,192 input symbols takes 144 ms
on NVIDIA V100 GPU, which degrades the throughput of compressing 1 GB data to less than 10 GB/s. 

Obviously, it is very important to develop an efficient parallel codebook construction algorithm on GPU to match the
high speed of other stages in Huffman encoding. To this end, we review the literature carefully for existing parallel
Huffman tree and codebook construction algorithms. Larmore and Przytycka \cite{larmore1995constructing} proposed a
parallel Huffman tree construction algorithm under the CREW PRAM model using $n$ processors with $\mathcal{O}(\sqrt{n}
\log n)$ time per processor, the first algorithm whose processor count scales linearly with the number of input symbols.
Here both the number of processors and the number of input symbols are $n$. However, the proposed algorithm is known for
its inefficiency of performing $\mathcal{O}(n^2)$ work. Millidi\'u et al. \cite{millidiu1999work} later proposed another
CREW PRAM algorithm for the same problem, with $n$ processors, $\mathcal{O}\big(H \cdot\log\log\frac{n}{H}\big)$ time
per processor, and $\mathcal{O}(n)$ work, where $H$ is the length of the longest codeword. Since Huffman codes can be up
to $\mathcal{O}(n)$ in length, the proposed algorithm has a worst case performance of $\mathcal{O}(n)$ per processor,
but this is rarely encountered in practice, especially in HPC scenarios whose floating-point data tends to be mostly
smooth and predictable.

The previously discussed algorithms all output Huffman trees, where the trees still need to be traversed serially to
generate a codebook. To address this issue, Ostadzadeh et al. proposed a CREW PRAM algorithm that directly generates the
codebook \cite{ostadzadeh2007practical}. To do this, it generates the length of each symbol, with $n$ processors and
$\mathcal{O}(H\!\cdot\!\log\log\frac{n}{H})$ time per processor, and then converts the generated symbol lengths into
codes, with $n$ processors and $\mathcal{O}(H)$ time per processor. We propose our Huffman codebook construction method
based on this algorithm due to its direct output of Huffman codes as well as its outstanding performance on most Huffman
work.

\section{Problem Statement}\label{sec:ps}

In this section, we first discuss the scalability constraints posed by modern GPU architectures, especially the CUDA
architecture. We then analyze the performance bottleneck of the state-of-the-art Huffman encoding method. Finally, we
formulate our research problem under certain constraints. 

\subsection{Scalability Constraints from GPU Hardware}\label{sub:ps-scale}

First, we analyze the scalability constraints from the GPU hardware perspective. The thread is the basic programmable
unit that allows the programmer to use the massive amount of CUDA cores. CUDA threads are grouped at different levels,
including warp, block, and grid levels. 


\paragraph{Rigid SIMD-ness Against Randomness} 
The \textit{warp} is a basic-level scheduling unit in CUDA associated with SIMD (single-instruction multiple-data).
Specifically, the threads in a warp achieve convergence when executing exactly the same instruction; otherwise, warp
divergence happens. In the current CUDA architecture, the number of threads in a warp is 32, hence, it works as 32-way
SIMT when converging. However, when diverging happens, it may cause discrepancy from the PRAM model, because diverged
threads add extra overhead to the execution. Thus, we relax the use of the PRAM model. Nevertheless, the GPU's massive
parallelism allows our implementation to exhibit the theoretical complexity of parallel Huffman coding under PRAM. 

\paragraph{Block-Shared Memory Lifecycle Binding}
Unlike the warp, the thread block (or simply block) is a less hardware-coupled description of thread organization, as it
is explicitly seen in the kernel configuration when launching one. Threads in the same block can access the shared
memory, a small pool of fast programmable cache. On one hand, shared memory is bound to active threads, which are
completely scheduled by the GPU hardware; however, on the other hand, a grid of threads may exceed the hardware
supported number of active threads at a time. As a result, the data stored in the shared memory used by the previous
batch of active threads may be invalid when the current or following batch of active threads are executing.

Therefore, we must make use of both coarse- and fine-grained parallelization in our design due to the scalability
constraints from the CUDA architecture. For coarse-grained parallelization, we divide the data into multiple independent
chunks, not only because it is easy to map chunks to thread blocks and utilize local shared memory, but also because it
will facilitate the reverse process, decoding. In addition, coarse-grained chunking can improve performance
significantly with only a minimum overhead in compression ratio. For fine-grained parallelization, the state-of-the-art
work \cite{Rahmani_Topal_Akinlar_2014} only addresses the encoding stage rather than Huffman codebook construction. In
the next section, we will further analyze the performance issue of the existing fine-grained encoding approach.

\subsection{Fully Enabling GPU's High Memory Bandwidth}
\label{sub:limit}

Compared to compute-bound algorithms such as matrix-matrix multiplication, Huffman encoding tends to be more
memory-bound \cite{habib2017balancing}. In practice, Huffman encoding that has not been highly optimized usually
underutilizes GPU memory bandwidth. In this section, we analyze the root causes of the existing method's low memory
bandwidth utilization, which will guide our design of an efficient Huffman encoding that fully enables high GPU memory
bandwidth.

\renewcommand{\arraystretch}{1.1}

\newcommand{\myCline}{\cline{2-4}\cline{6-7}\cline{9-11}}

\begin{table}[t]
\centering\small\sffamily

\caption{%
Parallelism implemented for Huffman coding's subprocedures (kernels). %
``sequential'' denotes that only 1 thread is used due to data dependency. %
``coarse-grained'' denotes that data is explicitly chunked. %
``fine-grained'' denotes that there is a data-thread mapping with little or no warp divergence.}%
\resizebox{\linewidth}{!}{\begin{tabular}{@{}r|r|r|c|r|r|c|r|r|r|r|l@{}}
\multicolumn{1}{r}{\BOLD {histogram}}
& \TabTitle{sequential}
& \TabTitle{coarse-grained}
& \TabTitle{fine-grained}
& \NullCell
& \TabTitleTL{{\color{gray}data-thread}}{many-to-one}
& \TabTitleTL{{\color{gray}data-thread}}{one-to-one}
& \NullCell
& \TabTitle{atomic write}
& \TabTitle{reduction}
& \TabTitle{prefix sum}
& boundary
\\  
\myCline
blockwise reduction
&
&
& \YES
&
& \multicolumn{1}{c|}{\YES}
&
&
& \multicolumn{1}{c|}{\YES}
& \multicolumn{1}{c|}{\YES}
&
& \SyncBlk
\\  
\myCline
gridwise reduction
&
&
& \YES
&
& \multicolumn{1}{c|}{\YES}
&
&
& \multicolumn{1}{c|}{\YES}
& \multicolumn{1}{c|}{\YES}
&
& \SyncDev
\\  
\myCline
\multicolumn{1}{r}{\BOLD{build codebook}}
& \NullCell
& \NullCell
& \NullCell
& \NullCell
& \NullCell
& \NullCell
& \NullCell
& \NullCell
& \NullCell
& \NullCell
& 
\\  
\myCline
get codeword lengths
&
& \YES
& \YES
&
& \YES
& \YES
& 
& \YES
&
&
& \SyncGrd
\\  
\myCline
get codewords
&
& 
& \YES
&
&
& \YES
&
& \YES
&
&
& \SyncGrd
\\  
\myCline
\multicolumn{1}{r}{\BOLD{canonize}}
& \NullCell
& \NullCell
& \NullCell
& \NullCell
& \NullCell
& \NullCell
& \NullCell
& \NullCell
& \NullCell
& \NullCell
& 
\\  
\myCline
get numl array
&
&
& \YES
&
&
& \YES
&
& \multicolumn{1}{c|}{\YES}
&
& \multicolumn{1}{c|}{\YES}
& \SyncGrd
\\  
\myCline
get first array (\textsc{raw})
& \multicolumn{1}{c|}{\YES}
&
&
&
& \multicolumn{1}{c|}{\YES}
&
&
&
&
&
& \SyncGrd
\\  
\myCline
canonization (\textsc{raw})
& \multicolumn{1}{c|}{\YES}
&
&
&
& \multicolumn{1}{c|}{\YES}
&
&
&
&
&
& \SyncGrd
\\  
\myCline
get reverse codebook
&
&
& \YES
&
&
&
&
&
&
&
& \SyncDev
\\  
\myCline
\multicolumn{1}{r}{\BOLD{Huffman enc.}}
& \NullCell
& \NullCell
& \NullCell
& \NullCell
& \NullCell
& \NullCell
& \NullCell
& \NullCell
& \NullCell
& \NullCell
& 
\\  
\myCline
\texttt{reduce}-merge
&
& \multicolumn{1}{c|}{\YES}
& \YES
&
& \multicolumn{1}{c|}{\YES}
&
&
&
& \multicolumn{1}{c|}{\YES}
&
& \SyncBlk
\\  
\myCline
\texttt{shuffle}-merge
&
& \multicolumn{1}{c|}{\YES}
& \YES
&
&
& \YES
&
&
&
&
& \SyncDev
\\  
\myCline
get blockwise code len
&
& \multicolumn{1}{c|}{\YES}
& \YES
&
&
& \YES
&
&
&
& \multicolumn{1}{c|}{\YES}
& \SyncGrd
\\  
\myCline
coalescing copy
&
& \YES
& \YES
&
&
& \YES
&
&
&
&
& \SyncDev
\\  
\myCline
\end{tabular}%
}
\label{tab:how-to-par}
\end{table}

\paragraph{Variable Lengths of Codewords}
Due to the variable lengths of Huffman codes, the serial encoding must calculate the location of each encoded symbol and
perform a \emph{write} operation. Thus, it is easy to implement a relatively efficient Huffman encoding on CPU because
of the CPU's sophisticated branch prediction and caching capabilities, which can effectively mitigate the irregular
memory access pattern, even with a relatively low memory bandwidth (e.g., Summit \cite{summit} has about a theoretical
peak memory bandwidth of about 60$\sim$135 GB/s). In comparison, the GPU has a much higher memory bandwidth but lower
branch prediction and caching capabilities. We note that coarse-grained parallel encoding (i.e., chunking data and
assigning each chunk to a processor) cannot fully utilize the GPU's high memory bandwidth, as it disregards memory
coalescing. This is confirmed by a prior work, \textsc{cuSZ} \cite{cusz2020} where coarse-grained parallel encoding only
achieves a throughput of about 30 GB/s on the V100 (1/30 of the peak).

\paragraph{Limitations of Existing GPU Encoding Method}
A prefix-sum based Huffman encoding algorithm was proposed to make use of the massive parallelism on GPUs
\cite{Rahmani_Topal_Akinlar_2014}. In this method, before memory copies, a classical parallel prefix-sum algorithm is
used to calculate the write locations of all encoded symbols. However, it has two main drawbacks to limit its use in all
scenarios. As discussed in Section~\ref{subs:bg-lossless}, many scientific applications generate the data that contains
the symbols each with more than one byte, thus, the lengths of the corresponding codewords are fairly variable and
diverse. On the one hand, the prefix-sum based method does not exhibit good performance in the high-compression-ratio
use cases (i.e., short codeword length averagely) \cite{cusz2020}. This is because, by moving only few bits in a
single-/multi-byte codeword, the codeword-length agnostic solution makes low use of the GPU memory bandwidth given the
same degree of launched parallelism, which is also confirmed by our experiment---the prefix-sum based method can only
achieve a throughput of 37 GB/s on V100 on a dataset with the average codeword length of 1.02717 bits. On the other
hand, even though the last step---concurrent write to global memory---is theoretically low in time complexity (i.e.,
$\mathcal{O}(1)$), the hardware implementation makes it tend to be CREW (exhibiting memory contention). For example, our
experiment shows that the concurrent iterative solution has similar performance as one-time exclusive parallel write. 

Overall, in this work, we aim to fully enable the high GPU bandwidth for Huffman encoding in a wide range of emerging
scenarios (i.e., more than 256 symbols in the codebook) without loss of generality. Note that achieving high performance
on GPUs requires to rigidly follow the coalescing and SIMD characteristics, which are against the irregular memory
access pattern. Therefore, in order to develop a high-performance Huffman encoder on GPUs, we need \Circled{1} to
balance the SIMD-ness from the GPU programming model and the inherent randomness of Huffman coding and \Circled{2} to
develop an adaptive solution to solve the low memory bandwidth utilization issue. 

\section{Design Methodology}
\label{sec:design}
In this section, we propose our novel GPU Huffman encoding design for the CUDA architecture. We propose several
optimizations for different stages. Specifically, we modularize the Huffman encoding into the following four stages:
\Circled{1} calculating the frequencies of all input symbols, namely, histogramming; \Circled{2} Huffman codebook
generation/construction based on the frequencies; \Circled{3} canonizing the codebook and generating the reverse
codebook for decoding; and \Circled{4} encoding according to the codebook, and concatenate Huffman codes into a
bitstream. We first propose an efficient, fine-grained parallel codebook construction on GPUs, especially for scenarios
requiring a large number of symbols (stage 2). We then propose a novel reduction-based encoding scheme that iteratively
merges the encoded symbols, which significantly improves memory bandwidth utilization (stage 4). A summary of our
proposed techniques is shown in Table 1. It shows the parallelism and CUDA APIs for each substage. We highlight the
corresponding \textit{granularity}, \textit{model}, \textit{scalability}, and \textit{complexity}. 

\subsection{Histogramming}
\label{sub:histo}

The first stage of Huffman encoding is to build a histogram representing the frequency of each integer-represented
symbol from the input data. The GPU histogramming algorithm in use is derived from that proposed by G\'omez-Luna et
al.~\cite{GmezLuna2012AnOA}. This algorithm minimizes conflicts in updating the histogram bin locations by replicating
the histogram for each thread block and storing the histogram in shared memory. Where possible, conflict is further
reduced by replicating the histogram such that each block has access to multiple copies. All threads inside a block read
a specified partition of the input and use atomic operations to update a specific replicated histogram. As each block
finishes its portion of the predicted data, the replicated histograms are combined via a parallel reduction into a
single global histogram, which is used to construct the final codebook. 

\subsection{Two-Phase Canonical Codebook Construction}
\label{sub:two-phase}
In the second stage, we implement an efficient parallel Huffman codebook construction algorithm on the GPU and modify it
to produce canonical codes for fast decoding. 

\subsubsection{Codebook Construction}\label{subs:construt}
Now that we have a single global histogram, the next step is to efficiently construct a base codebook. We implement the
parallel codebook construction algorithm proposed by Ostadzadeh et al., as described earlier, on the GPU
\cite{ostadzadeh2007practical}. This algorithm provides a parallel alternative to the original Huffman codebook
construction algorithm in $\mathcal{O}(n\log n)$ and directly generates codewords. To the extent of our knowledge,
this algorithm has not been implemented on the GPU elsewhere. The algorithm is split into two phases, \Circled{1}
\func{GenerateCL}, which calculates the codeword length for each input symbol, and \Circled{2} \func{GenerateCW}, which
generates the actual codeword for each input symbol. Both phases utilize fine-grain parallelism, with one thread mapped
to one input symbol or intermediate value. Additionally, both phases are implemented as single CUDA kernels with
Cooperative Groups \cite{cooperative-groups}, which we use to synchronize an entire CUDA grid. We describe both phases
in Algorithm~\ref{algo:par-build}, with our modifications colored blue, and emphasize our GPU implementation in the
following discussion. We refer readers to \cite{ostadzadeh2007practical} for more details of the original algorithm.

\begin{algorithm}
\centering
    \newcommand{\varhere}[1]{{\text{\fontfamily{lmr}\selectfont\textit{#1}}}}
\newcommand{\funchere}[1]{\textsc{\bfseries#1}}
\newcommand{\algoCommentFont}{\fontsize{6.5}{7}\selectfont\color{gray}\sffamily}
\newcommand{\IN}{\textcolor{R}{\bfseries in}}

\scriptsize\ttfamily
\begin{algorithmic}[1] 
\vspace{.5ex}
\Statex {\sffamily\color{R}$\bullet$\bfseries\small\ \func{GenerateCL} --- codeword length procedure}
\vspace{.5ex}
\State iNodes <- $\emptyset$, c <- 0
\ForAll{i {\IN} [0\ldots n) concurrently}
    \State lNodes[i].freq <- F[i], lNodes[i].leader <- (-1), CL[i] <- 0
\EndFor     \Comment{\textcolor{gray}{Initialize array of leaf nodes and set codeword lengths to zero}}
\While{c < n \textbf{\color{R}or} iNodes.size > 1}
    \State t <- \funchere{NewNodeFromSmallestTwo}({lNodes}, c, {iNodes}) 
    \State iNodes <- iNodes $\cup$ \{t\} \Comment{Create first internal node}
    \ForAll{i {\IN} [c\ldots n) concurrently}
        \If {lNodes[i].freq < t.freq}
            \State copy[i-c] <- lNodes[i]
            \State copy.size <- \funchere{AtomicMax}(i-c+1, copy.size)
        \EndIf
    \EndFor 
    \State c <- c + copy.size \Comment{Select eligible leaf nodes}
    \State $\ell$ <- copy.size + iNodes.size - 1             \hfill{\algoCommentFont of nodes $\looparrowleft$}
    \State s <- iNodes.size - 1 - ($\ell$ \textbf{mod} 2)    \Comment{Ensure \texttt{temp} will have an even number}
    \State temp <- \funchere{ParMerge}(copy, iNodes[0\ldots s) )   \Comment{Merge leaf and remainder of}
    \State iNodes <- iNodes[s\ldots iNodes.size]                   \hfill{\algoCommentFont internal nodes}
    \ForAll{i {\IN} [0\ldots temp.size/2) concurrently}
        \State iNodes[iNodes.size+i] <- \funchere{Meld}(temp[2$\cdot$i], temp[2$\cdot$i+1])
    \EndFor 
    \State iNodes.size <- iNodes.size + temp.size           \Comment{Meld each two adjacent nodes}
    \ForAll{i {\IN} [0\ldots n) concurrently}               \hfill{\algoCommentFont in parallel}
       \State \funchere{UpdateLeafNode}(lNode[i], CL[i])    \Comment{Update codeword lengths and leader}
    \EndFor                                                 \hfill{\algoCommentFont pointers for leaf nodes}
\EndWhile
\vspace{1ex}
\Statex {\sffamily\color{R}$\bullet$\bfseries\small\ \func{GenerateCW} --- codeword generation procedure}
\vspace{.5ex}
\State \funchere{ParReverse}(CL)
\State CCL <- CL[0], PCL <- CL[0], FCW <- the codeword 0, CDPI <- 0
\State First[CCL] <- 0, Entry[CCL] <- 0
\While{CDPI < n-1}
    \State new{CDPI} <- n-1
    \ForAll{i {\IN} [CDPI\ldots  n-1) concurrently}
        \If{CL[i] > CCL}
            \State new{CDPI} <- \funchere{AtomicMin}(new{CDPI}, i)
        \EndIf
    \EndFor \Comment {Count number of codewords with current codeword length}
    \color{B}
    \ForAll{i {\IN} [{CDPI}\ldots  new{CDPI}) concurrently}
        \State {CW[i] <- FCW + ({CDPI} - (newCDPI-i-1))}
    \EndFor \Comment{Build codewords of a given \texttt{CCL} in reverse}
    \State First[CCL] <- \funchere{InvertCW}(CW[newCDPI-1])
    \State Entry[CCL] <- Entry[PCL] + (newCDPI-CDPI) \Comment{Record decoding metadata}
    \State \color{black} CLDiff <- CL[newCDPI] - CL[newCDPI-1]
    \State FCW <- CW[CDPI] + 1 $\cdot\ $\funchere{pow}(2, CLDiff)
    \State PCL <- CCL, CCL <- CL[newCDPI], CDPI <- newCDPI  \Comment{Prepare for next}
\EndWhile \hfill{\algoCommentFont codeword length}\color{B}
\ForAll{i {\IN} [1\ldots n) concurrently}
    \State CW[i] <- \funchere{InvertCW}(CW[i])
\EndFor \Comment{Invert codewords, making them canonical and reordering them}
\color{black}
\State \funchere{ParReverse}(CW)
\end{algorithmic}

    \caption{Modified parallel Huffman code construction based on \cite{ostadzadeh2007practical}.}\label{algo:par-build}
\end{algorithm}

\func{GenerateCL} takes \var{F}, a sorted $n$-symbol histogram, and outputs \var{CL}, a size $n$ array of codeword lengths
for each symbol. This phase of the algorithm runs in $\mathcal{O}(H\cdot\log\log\frac{n}{H})$ time on PRAM, where $H$ is
the longest codeword. Its parallelism can be derived from the fact that for a given set of Huffman sub-trees, all
sub-trees whose total frequencies are less than the sum of the two smallest sub-tree frequencies can be combined in
parallel, a result which Ostadzadeh et al. prove \cite{ostadzadeh2007practical}. 

Before \func{GenerateCL} is launched, the histogram is sorted in ascending order using Thrust \cite{thrust}. This
operation is low-cost, as $n$ is relatively small compared to the input data size. Once launched, lines 1--4 of
Algorithm~\ref{algo:par-build} initialize the array \var{lNodes} with each input symbol's leaf node, and initialize the
array (and queue) \var{iNodes} as empty. Each array element describes a Huffman node, storing its total frequency, its
leader, or topmost parent, and auxiliary information. To increase memory access efficiency, each of these arrays are
stored in \textit{structure-of-arrays format}, rather than the intuitive array-of-structures format. The advantage of
this is that accesses to single fields of consecutive elements are coalesced. 

Next, lines 5--26 construct the Huffman tree while there are still leaf and internal nodes to process. Lines 6--16 create
a new node $t$ from the smallest two leaf or internal nodes, and selects leaf nodes whose frequencies are less than $t$.
\func{ParMerge} (line 17) merges these selected leaf nodes and internal nodes, which are sorted by ascending frequency,
together. This is done using a $\mathcal{O}(\log\log n)$ parallel merge under the PRAM model
\cite{ostadzadeh2007practical}.  

To implement this merge, we customize the parallel GPU Merge Path algorithm proposed by Green et al.
\cite{green2012mergepath} for our intermediate structure-of-arrays representations in \func{GenerateCL}. In practice,
this algorithm does not attain the proposed theoretical time complexity, as its time complexity is $\mathcal{O}(n / p +
\log n)$, with $p$ being the number of partitions (i.e., the number of thread blocks). However, we use a number of
thread blocks proportional to the number of streaming multiprocessors (SMs), making $n / p$ in practice
$\mathcal{O}(\log n)$. This component of parallel codebook construction employs coarse-grain parallelism, as once the
merge partitions are determined, each partition is merged serially. To remove the overhead of calling a separate kernel
with dynamic parallelism, we incorporate \func{ParMerge} into the same kernel as the rest of \func{GenerateCL}, keeping
unneeded threads idle until the merging phase. Once these nodes are merged, they are melded together into new nodes,
with the appropriate \var{CL} values and leaf nodes being updated, on lines 18--25, and the iteration is repeated as
appropriate. 

\func{GenerateCW} takes \var{CL} as input and outputs \var{CW} (i.e., the actual codewords). It takes $\mathcal{O}(H)$
time per thread in the PRAM model, where $H$ is the longest codeword, and our GPU implementation is consistent with this
theoretical complexity (see Table \ref{tab:construction-breakdown}). This phase of the algorithm utilizes fine-grained
parallelism, as codewords are generated by individual threads. It assigns numerically increasing codewords to all
symbols with a given codeword length \var{CCL} in parallel (lines 31--39). If there are more codewords to generate that
are longer than \var{CCL}, first, \var{CCL} and other local variables are updated (lines 42--44). Also, the first
codeword for the new codeword length is generated by incrementing the existing codeword and left-shifting the codeword
by the difference between the old and new codeword lengths (line 43). Finally, lines 31--39 are repeated. Once all
codewords are generated, \var{CW} is reversed, and the codewords are resorted by the actual input symbol they represent
to generate the forward codebook. 

It is worth noting that throughout codebook construction, we use the state-of-the-art CUDA Cooperative Groups instead of
existing block synchronization for global synchronization \cite{cooperative-groups}. This is because CUDA blocks are
limited to 1024 threads, and we employ fine-grained parallelism for codebook sizes greater than 1024. An alternative
technique to achieve the same global synchronization is to separate the parallel regions into different kernels. We
chose Cooperative Groups over this technique to avoid the overhead of kernel launches and \func{cudaDeviceSynchronize}.
Our profiling for the NVIDIA V100 GPU reveals that a CUDA kernel launch takes about 60 microseconds (60 $\mu$s), and
each of our parallel regions performs very little work. Also, since many of these parallel regions are performed in a
loop, we effectively avoid unnecessary CPU-GPU transfers. 

\subsubsection{Canonizing Codebook}\label{subs:canonize}
A canonical Huffman codebook~\cite{schwartz1964generating} holds the same bitwidth of each codeword as the original
Huffman codebook (i.e., base codebook). Its bijective mapping between input symbol and Huffman codeword is more
memory-efficient than Huffman-tree traverse for encoding/decoding. The time complexity of serially building a canonical
codebook from the base codebook is $\mathcal{O}(n)$, where $n$ is the number of symbols, and is sufficiently small
compared with the data size. By using a canonical codebook, we can \Circled{1} decode without the Huffman tree,
\Circled{2} efficiently cache the reverse codebook for high decoding throughput, and \Circled{3} maintain exactly the
same compression ratio as the base Huffman codebook. 

We start our implementation which contains a partially-parallelized canonization CUDA kernel, utilizing Cooperative
Groups. It performs \Circled{1} linear scanning of the base codebook (sequentially $\mathcal{O}(n)$), which is
parallelized at fine granularity with atomic operations; \Circled{2} loose radix-sorting of the codewords by bitwidth
(sequentially $\mathcal{O}(n)$), which cannot be parallelized because of the intrinsic RAW dependency; and \Circled{3}
building the reverse codebook (sequentially $\mathcal{O}(n)$), which is enabled with fine-grained parallelism. The
canonization process is relatively efficient---it only costs about $200$ us to canonize a $1024$-codeword codebook on V100.

Nevertheless, our choice of codebook construction algorithm provides a sufficient base for further optimization. Since
the output of \func{GenerateCL} and input of \func{GenerateCW}, \var{CL}, is sorted by codeword length, the intrinsic
RAW dependency of \Circled{2} is removed. In fact, the existing codewords generated by \func{GenerateCW} are almost
canonical, except for the fact that given two codewords $c_1$ and $c_2$ where $c_2$ is longer than $c_1$ and $\ell$ is
$c_1$'s length, the most significant $\ell$ bits of $c_2$ are numerically greater than $c_1$. The opposite should be the
case, as efficient decoding relies on this fact. To alleviate this problem, the codewords for each level are generated
in decreasing order per level (line 38 of Algorithm~\ref{algo:par-build}), and the bits in each codeword are then
inverted before they are stored (line 47). Moreover, additional metadata to facilitate decoding also needs to be
generated in $\mathcal{O}(1)$ extra time with little storage overhead. The metadata that we generate consists of two
$H$-element arrays, where $H$ is the longest codeword's length. The \var{First} array contains the first codeword for a
given length, and the \var{Entry} array contains a prefix sum of the number of codewords shorter than that length. Right
after all the codewords for a given length are generated, the \var{First} and \var{Entry} arrays are updated
appropriately at that index (lines 40--41). Both arrays are used for efficient treeless canonical decoding. 

The theoretical time complexity of our codebook construction, including our modifications to generate canonized codes,
is $\mathcal{O}(H\!\cdot\!\log\log\frac{n}{H})$; however, due to the particular implementation of merge, which is the
most expensive operation, the practical complexity is increased to $\mathcal{O}(H \cdot\frac{H}{n} / p + \log\frac{H}{n}
)$, where $H$ is the longest codeword length, $n$ is the number of symbols, and $p$ is the number of blocks launched.
Nevertheless, due to $H$ in practical being small and $p$ being sufficiently large, we observe the complexity of
$\mathcal{O}(\log n)$, and our experiments are consistent with this (see Table~\ref{tab:construction-breakdown}).

\subsection{Encoding}\label{subs:huffcoding}

We propose an iterative merge comprised of \func{reduce}-merge and \func{shuffle}-merge that is the key to improving
memory bandwidth utilization for encoding. In each iteration, every two codewords are merged into one, with their
lengths summed up. Formally, given two code-length tuples $(a,\ell)_{2k}$ and $(a,\ell)_{2k+1}$, we define 
\[
\text{\func{Merge}}\big( (a,\ell)_{2k}, (a,\ell)_{2k+1} \big) =
	\left(a_{2k}\oplus a_{2k+1}, \ell_{2k} + \ell_{2k+1}\right),
\]
where $\oplus$ represents for concatenating bits of $a_{2k+1}$ right after bits of $a_{2k}$. Note that the merge is not
\emph{commutative} for the encoded symbols and must follow the original order. 

We further split this merge into \func{reduce}-merge and \func{shuffle}-merge phases. Note that the first merge includes
a codebook lookup to get the codewords. After that, the merge is performed iteratively on two codewords each time. 

\paragraph{\func{Reduce}-Merge}\label{par:reduce-merge}
The practical average bitwidth of Huffman codewords can be fairly low (e.g., most are 1 or 2 bits in many HPC datasets),
while data movement is in terms of single-/multi-byte words (i.e. a multiple of 8 bits, such as
\texttt{uint\{8,16,32\}\_t}). This can significantly hurt the performance due to an extravagant use of threads on the
GPU if the data-thread mapping is too fine, e.g., 1-to-1. In each iteration of reduction, active threads for data
movement halve every iteration before bits saturates the representing words, leading to a waste of parallelism. Hence,
we map multiple codewords to one thread to merge until the merged bitwidth exceeds that a representing word can handle.
More precisely, the condition of stopping \func{reduce}-merge is that the average bitwidth of the merged codeword
exceeds half of the bitwidth of the data type. For example, merging codewords with an average bitwidth of 2.3 bits for 3
times is expected to result in an 8$\times$ length, averagely 18.4 bits, and end before going beyond 32 bits. The number
of iterations (denoted by \emph{reduction factor} $r$) can be determined by the entropy in bits of input data (from
histogram). After \func{reduce}-merge is done, the number of merged codewords is shrunk by $2^r\times$. 
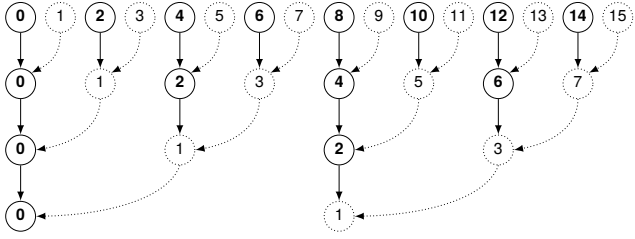
\begin{figure}
    \centering
\resizebox{0.96\columnwidth}{!}{%
    \begin{tikzpicture}[y=-1cm]\scriptsize

\tikzset{PairLeft/.style={draw, font=\strut\sffamily\bfseries, fill=none, circle, inner sep=0pt, minimum height=1.6\baselineskip, minimum width=1.6\baselineskip}};

\tikzset{PairRight/.style={draw, densely dotted, font=\strut\sffamily, fill=none, circle, inner sep=0pt, minimum height=1.6\baselineskip, minimum width=1.6\baselineskip}};

\begin{scope}[yscale=.5]

\def\loop{0}
\foreach \x/\content in 
{0/0, 1/1, 2/2, 3/3, 4/4, 5/5, 6/6, 7/7, 8/8, 9/9, 10/10, 11/11, 12/12, 13/13, 14/14, 15/15}
{

    \pgfmathparse{ Mod(\x, 2)==0 ? 1 : 0 }
    \ifnum\pgfmathresult > 0
    	\node[PairLeft] (\loop-\content) at (.6*\x, 0) {\content};
        
    \else
    	\node[PairRight] (\loop-\content) at (.6*\x, 0) {\content};
    \fi
}

\def\loop{1}
\foreach \x/\content/\text in 
{0/0/0, 1/1/2, 2/2/4, 3/3/6, 4/4/8, 5/5/10, 6/6/12, 7/7/14}
{
    \pgfmathparse{ Mod(\x, 2)==0 ? 1 : 0 }
    \ifnum\pgfmathresult > 0
    	\node[PairLeft] (\loop-\content) at (.6*\x*2, 2) {\content};
    \else
    	\node[PairRight] (\loop-\content) at (.6*\x*2, 2) {\content};
    \fi
}

\def\loop{2}
\foreach \x/\content/\text in 
{0/0/0, 1/1/4, 2/2/8, 3/3/12}
{
    \pgfmathparse{ Mod(\x, 2)==0 ? 1 : 0 }
    \ifnum\pgfmathresult > 0
    	\node[PairLeft] (\loop-\content) at (.6*\x*4, 4) {\content};
    \else
    	\node[PairRight] (\loop-\content) at (.6*\x*4, 4) {\content};
    \fi
}

\def\loop{3}
\foreach \x/\content\text in 
{0/0/0, 1/1/8}
{
    \pgfmathparse{ Mod(\x, 2)==0 ? 1 : 0 }
    \ifnum\pgfmathresult > 0
    	\node[PairLeft] (\loop-\content) at (.6*\x*8, 6) {\content};
    \else
    	\node[PairRight] (\loop-\content) at (.6*\x*8, 6) {\content};
    \fi
}

\foreach \x/\content in 
{0/0, 1/1, 2/2, 3/3, 4/4, 5/5, 6/6, 7/7, 8/8, 9/9, 10/10, 11/11, 12/12, 13/13, 14/14, 15/15}
{
    \pgfmathtruncatemacro\parity{Mod(\x,2)==0?1:0}
    \pgfmathtruncatemacro\next{int(\x/2)}
    \ifnum\parity > 0
        \draw[-latex] (0-\content.south) -- (1-\next.north); 
    \else
        \draw[-latex, densely dotted] (0-\content.south) to[out=-90, in=45] (1-\next.north east); 
    \fi
}

\foreach \x/\content in 
{0/0, 1/1, 2/2, 3/3, 4/4, 5/5, 6/6, 7/7}
{
    \pgfmathtruncatemacro\parity{Mod(\x,2)==0?1:0}
    \pgfmathtruncatemacro\next{int(\x/2)}
    \ifnum\parity > 0
        \draw[-latex] (1-\content.south) -- (2-\next.north); 
    \else
        \draw[-latex, densely dotted] (1-\content.south) to[out=-90, in=20] (2-\next.east); 
    \fi
}

\foreach \x/\content in 
{0/0, 1/1, 2/2, 3/3}
{
    \pgfmathtruncatemacro\parity{Mod(\x,2)==0?1:0}
    \pgfmathtruncatemacro\next{int(\x/2)}
    \ifnum\parity > 0
        \draw[-latex] (2-\content.south) -- (3-\next.north); 
    \else
        \draw[-latex, densely dotted] (2-\content.south) to[out=-100, in=2] (3-\next.east); 
    \fi
}

\end{scope}

\end{tikzpicture}
    }
    \caption{\func{reduce}-merge of 8-to-1.}
    \label{fig:reduce-merge}
\end{figure}

\begin{figure}
\centering
\includegraphics[%
    width=\linewidth, %
    trim=.1in 0 0 0
    ]{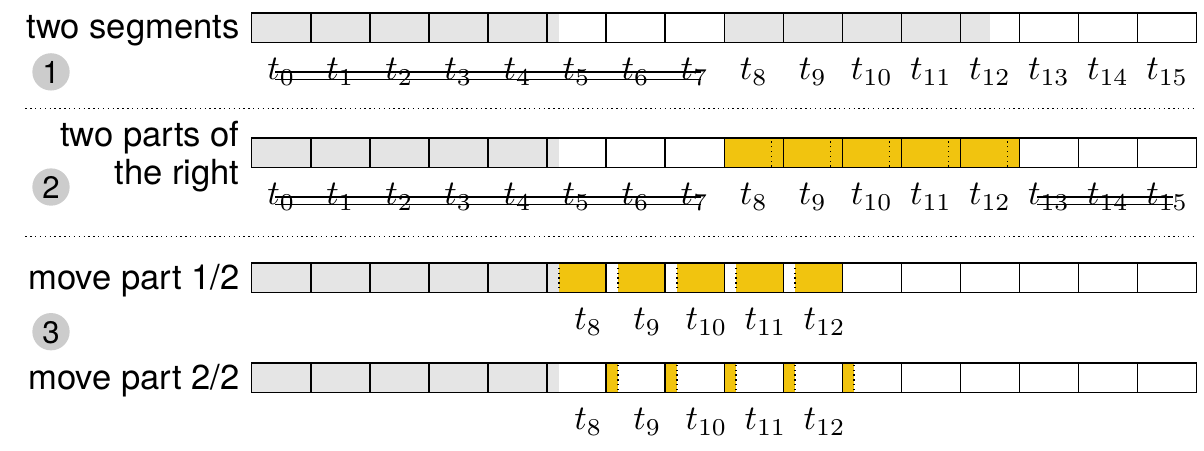}
\caption{Two-step batch move of grouped and typed data. By batch-moving the right grouped data, warp divergence is decreased.}
\label{fig:two-step}
\end{figure}

\paragraph{\func{Shuffle}-Merge}\label{par:shuffle-merge}
After \func{reduce}-merge completes, threads are grouped to move the corresponding merged codewords, with one thread
assigned to one unit of typed data. Following the same scheme as \func{reduce}-merge, a right group of typed data is
moved to append to its corresponding left group. To provide more detail, a left group $(A, \ell)_{2k}$, with data
segment (the representation data type or word is marked with $W$) and its length $\ell_W$, has a starting index $i_{2k}$
(already known) and an ending index $i_{2k,\bullet} \!=\! \big( i_{2k} +  \ell_{2k}/\ell_W \big)$ (easy to calculate).
Moreover, the ending bit's location can be calculated as $\ell_{2k,\bullet} \!=\! (\ell_{2k}\mod \ell_W)$, and the
number of residual bits is $\ell_{2k, \circ} \!=\! \ell_W \!-\! \ell_{2k,\bullet}$. Thus, the right group $(A,
\ell)_{2k+1}$ needs $\lceil\ell_{2k+1}/\ell_{H}\rceil$ threads for data movements. For each thread, the
$\ell_{2k,\circ}$ bits are first moved to fill the residual bits, and then the $\ell_{2k,\bullet}$ bits are moved right
after the $\ell_{2k,\circ}$ bits (in the next typed data cell), as shown in Figure~\ref{fig:two-step}. Note that this
process is free of data contention. The iterative process will be performed for $s$ times (denoted by \textit{shuffle
factor}) until a dense bitstream is formed. We also note that \func{shuffle}-merge can be finished within a continuous
memory space of $2^s$ typed data cells. 

\paragraph{Interface}

Our encoding kernel is interfaced as%
\begin{center}
    {\func{ReduceShuffleMerge}$\langle M,r\rangle$}(in, out, metadata).
\end{center}

We expose two independent parameters to describe the problem size and the two merge phases---magnitude $M$, reduction
factor $r$, and shuffle factor is derived from $s = M-r$. In total, there are $M$ iterations in the entire encoding
stage. After that, we generate a densely encoded bitstream for this chunk. We use $N \equiv 2^M$ to denote the problem
size for each chunk, and $n\equiv 2^s$ to denote the reduced size before \func{shuffle}-merge.  

We use the average codeword bitwidth to determine the reduction factor $r$. Specifically, given a word $W$ of length
$\ell_W$, a ``proper'' $r$ is tentatively determined according to $\lfloor \log \beta \rfloor \!+\! {r} \!+\! 1 \!=\!
\log \ell_W $, such that the length of the $r$-time-merged codeword $\ell^{(r)}$ is expected as $\ell_W/2 \le \ell^{(r)}
< \ell_W$, toward maximized memory bandwidth utilization. Note that there are certain codewords that exceed 32 bits
after $r$-time merge, and these (denoted by ``breaking'' in \TAB~\ref{tab:mag-rf-perf}) are filtered out and handled
otherwise. 

We give a quantitative example to illustrate the importance of determining a proper reduction factor and how magnitude
may affect the performance due to longer \texttt{shuffle}-merge. We compare a magnitude of $\{12, 11, 10\}$ and
reduction factor of $\{4,3,2\}$ on \texttt{Nyx-Quant} from \texttt{baryon-density} field, whose average bitwidth is 1.02717 with 1024 symbols. The
performance is shown in \TAB~\ref{tab:mag-rf-perf}. We find that the combination of $M$$=$$10$ and $r$$=$$3$ results in
the highest performance, which can be empirically generalized. Although reducing magnitude would result in more
metadata, as we prioritize the performance in this work, we choose the combination of $M\!=\!10$ and $r\!=\!3$ for the
following evaluation.




\begin{table}[ht]
   \TABFONTFAMILY
\caption{Performance (in GB/s) of our Huffman encoding with different chunk magnitudes (mag.) and reduction factors on Longhorn and Frontera.}
\label{tab:mag-rf-perf}
\resizebox{\linewidth}{!}{%
    \begin{tikzpicture}
    \node (mag) {
        \begin{tabular}{@{}r@{}}
            \TITLE mag. $\to$ \\
            \toprule
            (16$\times$) 4    \\
            (8$\times$) 3     \\
            (4$\times$) 2     \\
            \bottomrule
        \end{tabular}%
    };
    \node[right=0.25in of mag] (longhorn) {
        \begin{tabular}{@{}rrr@{}}
            \TITLE 2\textsuperscript{12} &
            \TITLE 2\textsuperscript{11} &
            \TITLE 2\textsuperscript{10}                   \\
            \toprule
            227.60                       & 274.40 & 291.04 \\
            191.41                       & 274.42 & 314.63 \\
            68.32                        & 106.87 & 172.54 \\
            \bottomrule
        \end{tabular}%
    };
    \node[right=0.25in of longhorn] (frontera) {
        \begin{tabular}{@{}rrr@{}}
            \TITLE 2\textsuperscript{12} &
            \TITLE 2\textsuperscript{11} &
            \TITLE 2\textsuperscript{10}                   \\
            \toprule
            110.94                       & 124.42 & 133.84 \\
            94.27                        & 124.56 & 135.86 \\
            42.70                        & 55.53  & 79.45  \\
            \bottomrule
        \end{tabular}%
    };
    \node[right=0.1in of frontera] (violating) {
        \begin{tabular}{@{}r@{}}
            \TITLE breaking \\
            \toprule
            0.000434\%      \\
            0.003277\%      \\
            0.007536\%      \\
            \bottomrule
        \end{tabular}%
    };
    \node[left=0.1in of mag.south west, anchor=west, rotate=90, text width=5em] (title1) {\large\BOLD reduction\\[-.5ex]factor};
    \node[left=0.05in of longhorn.south west, anchor=west, rotate=90] (title1) {\large\BOLD Longhorn};
    \node[left=0.05in of frontera.south west, anchor=west, rotate=90] (title2) {\large\BOLD Frontera};
\end{tikzpicture}
}
\end{table}

\begin{figure}[]
    \centering
    \includegraphics[trim={.1in 0 0 0}, width=\linewidth]{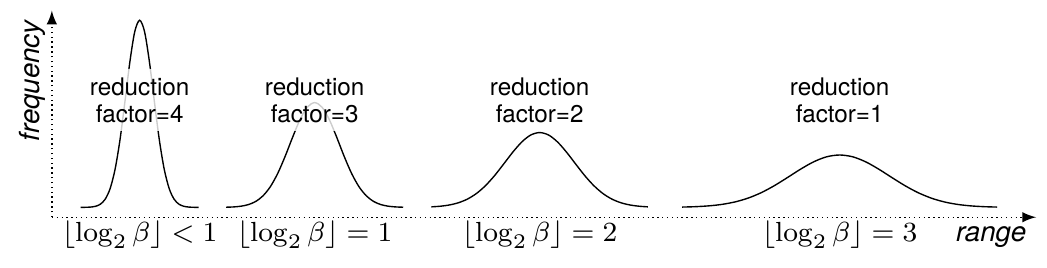}
    \caption{Average bitwidth being a consideration to decide reduction factor.}
    \label{fig:decide-rf}
\end{figure}

\paragraph{Complexity}
For \func{reduce}-merge, we map multiple codewords for reduction such that we can effectively move more bits against the
given holding codeword $W$ with length $\ell_W$. We maintain a block of $2^s$ threads for data movement. The time
complexity is $2^{(r - i)}$ for the $i$th iteration, and the time complexity for reduction in parallel is $\sum^r_1
2^{r-i}$. Note that the operations are homogeneous (i.e. there is only one if-branch) without being affected from warp divergence. 

For \func{shuffle}-merge, the magnitude of reduced data chunk remains $s$, and each typed data cell is assigned with a
thread. At the $i$th iteration, the chunk is split into $(s+1-i)$ groups. Compared with reduction, shuffle creates warp
divergence at a factor of 2, given the groups are to merge with their corresponding other groups. With $s$ parallel
\func{shuffle}-merges, the total time complexity is $\mathcal{O}(s)$. Note that bank conflicts may affect performance,
as the read and written locations inevitably overlap due to the variable length 
\footnote{Due to the page limit, we refer readers to the discussion of time complexity at 
\url{https://github.com/szcompressor/huffre/blob/main/doc/benchmark.md}.}. 
This proof-of-concept work is intended 
to show the effectiveness of the method, the effect of bank conflict is to be further investigated.

\section{Performance Evaluation}
\label{sec:eval}
In this section, we present our experimental setup (including platforms, baselines, and datasets) and our evaluation results.

\subsection{Experiment Setup}\label{sub:eval-setup}

\subsubsection{Evaluation Platforms}\label{subeval-setup-:platform}
We conduct our experimental evaluation using the Frontera supercomputer \cite{frontera} and its subsystem Longhorn
\cite{longhorn}. We perform our experiments on an NVIDIA Tesla V100 GPU from Longhorn and an NVIDIA Quadro RTX 5000 from
Frontera, and compare with CPU implementations on two 28-core Intel Xeon Platinum 8280 CPUs from Frontera.\footnote{V100
has 16GB HBM2 memory at 900 GB/s; RTX 5000 has 16GB GDDR6 memory at 448 GB/s; Xeon 8280 has 192GB of 2933 MT/s DDR4
memory.} We use NVIDIA CUDA 10.1 and its default profiler to measure the time. 
In this section, we use {\gpuRTX}to denote Turing RTX 5000 and {\gpuVolta}to denote Volta V100.

\subsubsection{Comparison Baselines}\label{sub:eval-setup-baseline}
CUHD \cite{weissenberger2018massively} and \textsc{cuSZ} \cite{cusz2020} are two state-of-the-art Huffman encoders for
GPUs, but both of them are coarse-grained and embarrassingly parallelized. We note that CUHD's source code
\cite{cuhd-github} only focuses on GPU Huffman decoding and implements a serial CPU Huffman encoder, so we compare our
GPU Huffman encoder only with \textsc{cuSZ} \cite{cusz2020}. We also compare our GPU encoder with the serial encoder
implemented in SZ \cite{sz-github} and with our implemented multi-thread encoder\footnote{Note that SZ's current OpenMP
version \cite{sz-omp} only divides data into multiple blocks and applies its compression to each block independently.}
on single and multiple CPU cores, respectively.

\subsubsection{Test Datasets}\label{subs:eval-setup-dataset}

\paragraph{Single-Byte Based Datasets}
Generic Huffman coding takes one byte per symbol, hence, at most 256 symbols in total. Without reinterpreting a
bytestream into multi-byte (un)signed integers or floating-point numbers, the generic encoding simply treats all input
data as \verb+uint8_t+. Our evaluation includes these popular datasets: \Circled{1} \texttt{enwik8} and \texttt{enwik9}
from \textit{Large Text Compression Benchmark} \cite{ltcb}, the first $10^8$ and $10^9$ bytes of XML-based English
Wikipedia dump; \Circled{2} \texttt{nci} from \textit{Silesia Corpus} \cite{silesia}, a file for chemical database of
structures; \Circled{3} \texttt{mr} from \textit{Silesia Corpus} \cite{silesia}, a sample file of medical magnetic
resonance image; and \Circled{4} \texttt{Flan\_1565} from \textit{SuiteSparse Matrix Collection} \cite{ssmc}, a sparse
matrix in Rutherford Boeing format.  

\paragraph{Multi-Byte Based Datasets}

We also evaluate two datasets with multiple bytes as a symbol: \Circled{5} \texttt {Nyx-Quant} is the quantization codes
generated by SZ (a famous error-bounded lossy compression for HPC data) based on \textit{Nyx}'s (cosmological
simulation) \texttt{baryon\_density} from \textit{Scientific Data Reduction Benchmarks} \cite{sdrbench}; and \Circled{6}
\texttt{gbbct1.seq} is a sample DNA sequence data from \textit{GenBank} \cite{benson2012genbank}, where every $k$
nucleotides ($k$-mer) forms a symbol. We test $k=\{3,4,5\}$ in our evaluation. 

\subsection{Experimental Results}\label{sub:eval-result}

\subsubsection{Parallel Codebook Construction}\label{subs:eval-result-tree}

Call back to \SEC\ref{subs:construt}, we observe a practical complexity of $\mathcal{O}(H\cdot \log(n / H))$ or
approximately $\mathcal{O}(\log n)$, where $H$ and $n$ are the height of the built tree and the problem size,
respectively. Although we exhibit speedups in codebook construction for $n \!=\! 256$ ranging from 2.0$\sim$2.9 in
Table~\ref{tab:breakdown-kernel}, greater benefits come from using more input symbols since the $\mathcal{O}(n\cdot\log
n)$ serial construction algorithm scales slower. To demonstrate this speedup, we use the \texttt{gbbct1.seq} gene
dataset with $k$-mer analytics, where $k \!=\! 3, 4, 5$. We also evaluate our codebook construction on the quantization
codes generated by \textsc{cuSZ} from the Nyx dataset. Note that data other than the 4 bases of DNA are stored in
\texttt{gbbct1.seq}, and as a result, the number of input symbols needed is greater than $4^k$. 

\begin{table}[ht]
    \TABFONTFAMILY
    \caption{Breakdown comparison of Huffman codebook construction time (in milliseconds) on RTX 5000 and V100 with different numbers of symbols. }
    \resizebox{\linewidth}{!}{\begin{tikzpicture}

    \node[anchor=west]{%
        \begin{tabular}{@{} l @{}r @{}r rr rr rr @{}}
                      &
                      & \TITLE ref. CPU                      & \gpuRTX & \gpuVolta & \gpuRTX & \gpuVolta & \gpuRTX & \gpuVolta \\
                      & \EvalTableTitle{1}{r@{}}{\#sym.}
                      & \EvalTableTitle{1}{r}{serial}
                      & \EvalTableTitle{2}{r}{gen. codebook}
                      & \EvalTableTitle{2}{r}{canonize}
                      & \EvalTableTitle{2}{r@{}}{total time}                                                                   \\
            \cmidrule[.75pt](){1-2}\cmidrule[.75pt](lr){3-3}\cmidrule[.75pt](lr){4-5}\cmidrule[.75pt](lr){6-7}\cmidrule[.75pt](l){8-9}
            Nyx-Quant & 1024                                                                                                   
                      & 0.045                                & 3.051   & 3.689     & 0.095   & 0.115     & 3.416   & 3.804     \\
            {3-mer}   & 2048                                                                                                   
                      & 0.208                                & 8.381   & 9.760     & 0.242   & 0.284     & 8.623   & 10.044    \\
            {4-mer}   & 4096                                                                                                   
                      & 0.695                                & 20.148  & 24.684    & 0.519   & 0.663     & 20.667  & 25.347    \\
            {5-mer}   & 8192                                                                                                   
                      & 1.806                                & 61.748  & 59.092    & 1.453   & 1.449     & 63.201  & 60.541    \\[2ex]
                      & \EvalTableTitle{1}{r@{}}{\#sym.}
                      & \EvalTableTitle{1}{r}{serial}
                      & \EvalTableTitle{2}{r}{gen. CL}
                      & \EvalTableTitle{2}{r}{gen. CW}
                      & \EvalTableTitle{2}{r@{}}{total time}                                                                   \\
            \cmidrule[.75pt](){1-2}\cmidrule[.75pt](lr){3-3}\cmidrule[.75pt](lr){4-5}\cmidrule[.75pt](lr){6-7}\cmidrule[.75pt](l){8-9}
            Nyx-Quant & 1024                                                                                                   
                      & 0.045                                & 0.315   & 0.383     & 0.134   & 0.161     & 0.449   & 0.544     \\
            {3-mer}   & 2048                                                                                                   
                      & 0.208                                & 0.494   & 0.570     & 0.180   & 0.209     & 0.674   & 0.779     \\
            {4-mer}   & 4096                                                                                                   
                      & 0.695                                & 0.633   & 0.682     & 0.173   & 0.185     & 0.806   & 0.867     \\
            {5-mer}   & 8192                                                                                                   
                      & 1.806                                & 1.330   & 1.145     & 0.154   & 0.187     & 1.484   & 1.332     \\
        \end{tabular}
    };
    \node[rotate=90, anchor=west, xshift=-.03in, yshift=.2in]{{\large\BOLD cuSZ} serial};

    \node[rotate=90, anchor=west, xshift=-1.15in, yshift=.2in]{{\large\BOLD Ours} parallel};
\end{tikzpicture}}
    \label{tab:construction-breakdown}
\end{table}

Table~\ref{tab:construction-breakdown} compares GPU codebook construction between \textsc{cuSZ}'s serial implementation
and our parallel implementation on several datasets with different numbers of input symbols. Ours exhibits more dramatic
speedups over \textsc{cuSZ}'s when using more input symbols, consistent with our theoretical analysis and performing up
to 45.5$\times$ faster when creating a codebook for $8192$ symbols. Note that ours is no faster than the CPU serial
construction when the number of symbols is below $8192$. This is because caching, high frequency, and superior branch
prediction in combination result in low latency of CPU threads. However, to avoid long histogramming and other CPU-GPU
data transfers, it is desirable to purely perform codebook construction on the GPU. 

\begin{table}[ht]
    \TABFONTFAMILY
    \caption{Performance (in milliseconds) of multi-thread codebook construction with different numbers of input symbols. The length of the bar under the number reflects the execution time.}
    \centering
    \resizebox{\linewidth}{!}{%
        \newcommand{\BAR}[1]{\begin{tikzpicture}\draw[draw=none, fill=black!80] (0,0) rectangle ++({#1*0.4}, 0.08);\end{tikzpicture}}
\newcommand{\SUPERBOLD}{\fontfamily{ugq}\selectfont}

\begin{tabular}{@{}l r *{6}{r} @{}}
                   &
    \TITLE \#sym.  &
    \TITLE serial  &
    \TITLE 1 core  &
    \TITLE 2 cores &
    \TITLE 4 cores &
    \TITLE 6 cores &
    \TITLE 8 cores                                                                                                                  \\
    \cmidrule[.75pt](r){1-2}
    \cmidrule[.75pt](lr){3-3}
    \cmidrule[.75pt](l){4-8}
    Nyx-Quant      & 1024  & \SUPERBOLD 0.045 & 0.219            & 0.469         & 0.622            & 0.700         & 0.840         \\[-1.9ex]
                   &       & \BAR{0.30103}    & \BAR{0.76839}    & \BAR{1.05775} & \BAR{1.17091}    & \BAR{1.21894} & \BAR{1.29373} \\
    3-mer          & 2048  & \SUPERBOLD 0.208 & 0.361            & 0.691         & 1.101            & 1.122         & 1.303         \\[-1.9ex]
                   &       & \BAR{0.74991}    & \BAR{0.95531}    & \BAR{1.21367} & \BAR{1.40597}    & \BAR{1.41386} & \BAR{1.47648} \\
    4-mer          & 4096  & 0.695            & \SUPERBOLD 0.626 & 1.006         & 1.309            & 1.456         & 1.707         \\[-1.9ex]
                   &       & \BAR{1.21602}    & \BAR{1.17351}    & \BAR{1.36839} & \BAR{1.47841}    & \BAR{1.52317} & \BAR{1.59032} \\
    5-mer          & 8192  & 1.806            & \SUPERBOLD 1.167 & 1.513         & 1.657            & 1.836         & 2.158         \\[-1.9ex]
                   &       & \BAR{1.61419}    & \BAR{1.43029}    & \BAR{1.53935} & \BAR{1.57775}    & \BAR{1.62118} & \BAR{1.68980} \\
    Synthetic      & 16384 & 3.671            & \SUPERBOLD 1.683 & 1.796         & 1.705            & 2.055         & 2.222         \\[-1.9ex]
                   &       & \BAR{1.91686}    & \BAR{1.58433}    & \BAR{1.61184} & \BAR{1.58983}    & \BAR{1.66901} & \BAR{1.70224} \\
    Synthetic      & 32768 & 5.783            & 2.974            & 2.858         & \SUPERBOLD 2.626 & 2.873         & 3.139         \\[-1.9ex]
                   &       & \BAR{2.11231}    & \BAR{1.82665}    & \BAR{1.80963} & \BAR{1.77346}    & \BAR{1.81187} & \BAR{1.84976} \\
    Synthetic      & 65536 & 7.641            & 5.221            & 4.850         & \SUPERBOLD 4.411 & 4.952         & 5.713         \\[-1.9ex]
                   &       & \BAR{2.23249}    & \BAR{2.06827}    & \BAR{2.03654} & \BAR{1.99573}    & \BAR{2.04550} & \BAR{2.10706} \\
\end{tabular}%

    }
    \label{tab:omp-par}
\end{table}

\begin{table*}[ht]
    \TABFONTFAMILY
    \centering\small
    \caption{Breakdown comparison of Huffman performance on tested datasets. Gathering time is excluded.}
    \resizebox{0.8\linewidth}{!}{
\begin{tabular}{@{} l@{}rl cc c rr rr rr rr @{}}
     &
     &
     &
     &
     &
     & \gpuRTX
     & \gpuVolta
     & \gpuRTX
     & \gpuVolta
     & \gpuRTX
     & \gpuVolta
     & \gpuRTX
     & \gpuVolta
    \\
     &
     &
     & \TITLE avg. bits
     & \TITLE \#reduce
     & \TITLE breaking
     & \EvalTableTitle{2}{r}{hist.\enspace\Unit{gb/s}}
     & \EvalTableTitle{2}{r}{codebook\enspace\Unit{ms}}
     & \EvalTableTitle{2}{r}{encode\enspace\Unit{gb/s}}
     & \EvalTableTitle{2}{r@{}}{hist+enc\enspace\Unit{gb/s}}
    \\
    \cmidrule[.75pt](r){1-3}
    \cmidrule[.75pt](lr){4-5}
    \cmidrule[.75pt](lr){6-6}
    \cmidrule[.75pt](lr){7-8}
    \cmidrule[.75pt](lr){9-10}
    \cmidrule[.75pt](lr){11-12}
    \cmidrule[.75pt](l){13-14}
    {\sffamily\bfseries\scshape cuSZ}
     & {enwik8}
     & \DatumSize{95}{mb}
     & 5.1639                                               
     & \multicolumn{1}{c}{-}                                
     & \multicolumn{1}{@{}c}{-} 
     & 102.5
     & 252.4
     & 1.375
     & 1.635
     & 10.1                                                 
     & 12.2                                                 
     & 8.2
     & 9.8
    \\
     & {enwik9}
     & \DatumSize{954}{mb}
     & 5.2124                                               
     & \multicolumn{1}{c}{-}                                
     & \multicolumn{1}{@{}c}{-} 
     & 108.2
     & 259.6
     & 1.382
     & 1.640
     & 7.2                                                  
     & 11.3                                                 
     & 6.8
     & 10.8
    \\
     & {mr}
     & \DatumSize{9.5}{mb}
     & 4.0165                                               
     & \multicolumn{1}{c}{-}                                
     & \multicolumn{1}{@{}c}{-} 
     & 36.2
     & 86.5
     & 1.565
     & 1.831
     & 9.6                                                  
     & 15.2                                                 
     & 3.5
     & 3.8
    \\
     & {nci}
     & \DatumSize{32}{mb}
     & 2.7307                                               
     & \multicolumn{1}{c}{-}                                
     & \multicolumn{1}{@{}c}{-} 
     & 66.1
     & 150.6
     & 0.706
     & 1.027
     & 8.6                                                  
     & 14.9                                                 
     & 6.6
     & 9.6
    \\
     & {Flan\_1565}
     & \DatumSize{1.4}{gb}
     & 4.1428                                               
     & \multicolumn{1}{c}{-}                                
     & \multicolumn{1}{@{}c}{-} 
     & 104.2
     & 256.6
     & 0.758
     & 0.950
     & 8.5                                                  
     & 10.7                                                 
     & 7.8
     & 10.2
    \\
     & {Nyx-Quant}
     & \DatumSize{256}{mb}
     & 1.0272                                               
     & \multicolumn{1}{c}{-}                                
     & \multicolumn{1}{@{}c}{-} 
     & 74.8                                                 
     & 197.7                                                
     & 3.416                                                
     & 3.804                                                
     & 17.7                                                 
     & 29.7                                                 
     & 12.1
     & 18.9
    \\
    \cmidrule(r){1-3}
    \cmidrule(lr){4-5}
    \cmidrule(lr){6-6}
    \cmidrule(lr){7-8}
    \cmidrule(lr){9-10}
    \cmidrule(lr){11-12}
    \cmidrule(l){13-14}
    {\sffamily\bfseries Ours}
     & {enwik8}
     & \DatumSize{95}{mb}
     & 5.1639                                               
     & 2 (4$\times$)                                        
     & 0.034915\%                                           
     & 102.8
     & 252.0
     & 0.594
     & 0.707
     & 42.2                                                 
     & 94.0                                                 
     & 25.4
     & 46.1
    \\
     & {enwik9}
     & \DatumSize{954}{mb}
     & 5.2124                                               
     & 2 (4$\times$)                                        
     & 0.021747\%                                           
     & 108.1
     & 276.1
     & 0.626
     & 0.666
     & 49.7                                                 
     & 94.6                                                 
     & 34.0
     & 70.6
    \\
     & {mr}
     & \DatumSize{9.5}{mb}
     & 4.0165                                               
     & 2 (4$\times$)                                        
     & 0.000174\%                                           
     & 36.2
     & 99.0
     & 0.300
     & 0.312
     & 42.0                                                 
     & 76.8                                                 
     & 12.3
     & 18.4
    \\
     & {nci}
     & \DatumSize{32}{mb}
     & 2.7307                                               
     & 3 (8$\times$)                                        
     & 0.152880\%                                           
     & 56.4
     & 169.1
     & 0.507
     & 0.514
     & 63.7                                                 
     & 154.8                                                
     & 20.6
     & 36.1
    \\
     & {Flan\_1565}
     & \DatumSize{1.4}{gb}
     & 4.1428                                               
     & 2 (4$\times$)                                        
     & nearly 0\%                                           
     & 103.5
     & 274.7
     & 0.314
     & 0.327
     & 50.0                                                 
     & 94.9                                                 
     & 33.5
     & 69.5
    \\
     & {Nyx-Quant}
     & \DatumSize{256}{mb}
     & 1.0272                                               
     & 3 (8$\times$)                                        
     & 0.003277\%                                           
     & 74.8                                                 
     & 197.6                                                
     & 0.449                                                
     & 0.544                                                
     & 145.2                                                
     & 314.6                                                
     & 45.4
     & 96.0
    \\
    \cmidrule[.75pt](r){1-3}
    \cmidrule[.75pt](lr){4-5}
    \cmidrule[.75pt](lr){6-6}
    \cmidrule[.75pt](lr){7-8}
    \cmidrule[.75pt](lr){9-10}
    \cmidrule[.75pt](lr){11-12}
    \cmidrule[.75pt](l){13-14}
\end{tabular}

}
    \label{tab:breakdown-kernel}
    \vspace{-4mm}
\end{table*}

Moreover, since SZ \cite{sz-github} currently does not support building the Huffman tree in parallel, we also implement
a multi-thread codebook construction using OpenMP. We evaluate its performance and compare it with SZ's serial codebook
construction on our tested datasets with $1024\sim 8192$ symbols and synthetic normally-distributed
histograms\footnote{The symbol numbers in the tested real datasets are no more than $8192$, so we use synthetic data for
more than $8192$ symbols. Also, note that $8192$ is limited by the current optimal GPU histogramming.} with
16384$\sim$65536 symbols, as shown in \TAB~\ref{tab:omp-par}. We note that in many cases, even with only one thread, its
performance is better than the serial construction, as it uses internal cache-friendly arrays rather than the binary
trees and priority queues used by the serial construction. We also note that on our test datasets (with codebooks in the
order of $10^3$), the multi-thread construction does not improve the performance because the OpenMP introduces more
overhead than it reduces from multiple threads.  

We find that our multi-thread CPU codebook construction needs least $32768$ symbols to be able to overcome the OpenMP
overhead and obtain a speedup using multiple threads. Unlike CPU, GPU parallel construction can always yield a speedup
over serial construction in our tested cases. This is due to the relatively high latency and low performance of a single
GPU thread. On the other hand, parallelization and synchronization are relatively low-latency on the GPU. Furthermore,
since the GPU can launch vastly more threads than the CPU, it takes advantage of fine-grained parallelism in our
parallel codebook construction algorithm. 

\subsubsection{Encoding}\label{subs:eval-result-enc}

Without loss of generality, we evaluate our Huffman encoder on multiple datasets with various types, as shown in
\TAB~\ref{tab:breakdown-kernel}. Most of our tested datasets have a relatively large average bitwidth (e.g. 4 or more
bits vs. uncompressed 8 bits), leading to a relatively low compression ratio. According to the aforementioned $r$
decision-making mechanism, only \texttt{nci} and \texttt{Nyx-Quant} can use $r$$=$$3$ (potentially $r$$=$$4$ for
\texttt{Nyx-Quant}), making their throughputs over 100 GB/s. While compared with \texttt{Nyx-Quant}, \texttt{nci} has a
relatively small data size, so it is difficult to undersaturate the memory bandwidth. Moreover, \texttt{Nyx-Quant} has
much lower writing effort due to 2.66$\times$ higher compression ratio than \texttt{nci}, so its encoding throughput is high, at
314.6 GB/s.

As mentioned in \SEC\ref{subs:huffcoding}, the rigid fixed-size typed data makes it possible to meet conflict. For
example, when merging \verb+Flan_1565+, there is less than 1.4e-6\% of the data that breaks the fixed size, while there
is 3e-3\% of the data for \texttt{Nyx-Quant}. Our solution is to backtrace the breaking points in batch, which starts at the last iteration of \texttt{reduce}-merge and refers to the $2^r$ points that go beyond
32-bit limit together.
The reduction is about 300 $\mu$s, including one-time read from global memory.
The total number of breaking points (in percentile) are shown in the  ``breaking'' column in
\TAB~\ref{tab:breakdown-kernel}, which is negligible to affect the compression ratio. After we filter out the breaking
data, we can use the state-of-the-art \texttt{cuSPARSE} API or perform a ``device-wide'' using ready tools such as \texttt{NVIDIA::cub} \cite{nvidia-cub} to perform a dense-to-sparse conversion to save them.
In addition, optional gathering coarse-grained chunks into an even denser format can be done, using the merged length; 
due to the compression ratio (CR), there is only as much as 1/CR additional data movement, which is a short-time \texttt{memcpy} to the \texttt{Reduce-Shuffle-}merge time.
Compared with \textsc{cuSZ}'s implementation, our encoder can improve the performance by
$3.1\times\sim5.0\times$ and $3.8\times\sim6.8\times$ on RTX 5000 and V100, respectively, on the tested data.

\begin{table}
\centering\TABFONTFAMILY
\caption{Performance of multi-thread Huffman encoder on \texttt{Nyx-Quant}.}
\resizebox{\linewidth}{!}{%
    \begin{tabular}{@{}l rrrrrrrr@{}}
    \TITLE cores                & \TITLE 1                 & \TITLE 2         & \TITLE 4         & \TITLE 8         & \TITLE 16        & \TITLE 32        & \fontfamily{ugq}\selectfont56 & \TITLE\color{gray}64 \\
    \toprule
    \TITLE hist. (GB/s)         & 2.24                     & 4.42             & 8.83             & 17.61            & 34.97            & 63.59            & 61.47                         & \color{gray}63.14    \\
    \color{gray}par. efficiency & \color{gray}1.00         & \color{gray}0.99 & \color{gray}0.98 & \color{gray}0.98 & \color{gray}0.97 & \color{gray}0.89 & \color{gray}0.49              & \color{gray}0.44     \\
    \midrule
    \TITLE codebook (ms)        & \multicolumn{8}{c}{0.22}                                                                                                                                                       \\
    \midrule
    \TITLE enc. (GB/s)          & 1.22                     & 2.43             & 4.83             & 9.64             & 19.16            & 37.85            & 55.71                         & \color{gray}29.33    \\
    \color{gray}par. efficiency & \color{gray}1.00         & \color{gray}0.99 & \color{gray}0.99 & \color{gray}0.99 & \color{gray}0.98 & \color{gray}0.97 & \color{gray}0.81              & \color{gray}0.37     \\
    \midrule
    \TITLE hist+enc (GB/s)       & 0.79                     & 1.57             & 3.12             & 6.23             & 12.38            & 23.73            & 29.22                         & \color{gray}20.03    \\
    \bottomrule
\end{tabular}%
\enspace
\begin{tabular}{@{}rr@{}}
    \gpuRTX & \gpuVolta \\
    \toprule
    74.80   & 197.60    \\
            &           \\
    \midrule
    0.45    & 0.54      \\
    \midrule
    145.20  & 314.60    \\
            &           \\
    \midrule
    45.35   & 96.01     \\
    \bottomrule
\end{tabular}%
}
\label{tab:omp-huffman}%
\end{table}%

Finally, we evaluate the performance of our implemented multi-thread encoding and our overall encoder on
\texttt{Nyx-Quant}\footnote{We note that multi-thread histogramming/encoding have relatively stable throughput, thus we
only evaluate on \texttt{Nyx-Quant} for demonstration purpose.} with different numbers of CPU cores, as shown in
Table~\ref{tab:omp-huffman}. For encoding, the multi-thread version achieves a peak performance of    56 GB/s, and
maintains high parallel efficiency up to 32 cores (with 56 available cores). However, this is still about $5.6\times$
lower than the performance of our fine-grained Huffman encoding on the V100 (i.e., 314.6 GB/s). For the overall encoder
(including histogramming, codebook construction, and encoding), compared with the multi-thread encoder on the CPUs, our
GPU version on the V100 improves the histogram-encoding performance by 3.3$\times$ (i.e., 29.22 GB/s v.s. 96.01 GB/s). This is because
Huffman encoding tends to be memory-bound, and the GPU memory such as HBM2 in the V100 has a much higher bandwidth than
state-of-the-art CPU memory such as DRAM4.

\section{Related Work}
\label{sec:related}

We note that several approaches \cite{angulo2015accelerating,weissenberger2018massively} have been proposed to
accelerate Huffman decoding on GPUs, yet few studies considered optimizing Huffman encoding by fully utilizing GPU
computing power. Despite the limited work on GPU-based Huffman encoding, we still search for some related works and
discuss them as follows. 

In general, parallel Huffman coding obtains each codeword from a lookup table (generated by a Huffman tree) and
concatenates codewords together with other codewords. However, a severe performance issue arises when different threads
write codewords of varying lengths, which results in warp divergence on GPU \cite{xiang2014warp}. The most deviation
between methods occurs in concatenating codewords. Fuentes-Alventosa et al.
\cite{Fuentes-Alventosa_Gomez-Luna_Gonzalez-Linares_Guil_2014} proposed a CUDA implementation of Huffman coding with a
given table of variable-length codes, reaching 20$\times$ the serial CPU encoding performance. Rahmani et al.
\cite{Rahmani_Topal_Akinlar_2014} also proposed a CUDA implementation of Huffman coding based on serially constructing
the Huffman codeword tree and parallel generating the bytestream, which can achieve up to 22$\times$ over the serial CPU
performance by disregarding constraint on the maximum codeword length or data entropy. Lal et al.
\cite{Lal_Lucas_Juurlink_2017} proposed a Huffman-coding-based memory compression technique for GPUs (called
E\textsuperscript{2}MC) based on a probability estimation of symbols. Recently, Tian et al. \cite{cusz2020} developed a
coarse-grained parallel Huffman encoder for error-bounded lossy compressor on GPUs; however, this implementation does
not address the non-coalesced memory issue, reaching only 30 GB/s on V100. 

\section{Conclusion and Future Work}
\label{sec:conclusion}
In this work, we propose and implement an efficient Huffman encoder for NVIDIA GPU architectures. Specifically, we
develop an efficient parallel codebook construction and a novel reduction based encoding scheme for GPUs. We also
implement a multi-thread Huffman encoder for a fair comparison. We evaluate our encoder using six real-world datasets on
NVIDIA RTX 5000 and V100 GPUs. Compared with the state-of-the-art Huffman encoder, our solution can improve the parallel
encoding performance up to 5.0$\times$ on RTX 5000, 6.8$\times$ on V100, and 3.3$\times$ on CPUs. We plan to further
optimize the performance in the future work. For example, 
we seek to (1) tune the performance for low-compression-ratio data, 
(2) explore more efficient gathering methods,
and (3) explore how intrinsic data feature affects the compression ratio and the throughput. 


\section*{Acknowledgements}
\label{sec:acknowledgement}
\small This research was supported by the Exascale Computing Project (ECP), Project Number: 17-SC-20-SC, a collaborative
effort of two DOE organizations---the Office of Science and the National Nuclear Security Administration, responsible
for the planning and preparation of a capable exascale ecosystem, including software, applications, hardware, advanced
system engineering and early testbed platforms, to support the nation's exascale computing imperative. The material was
supported by the U.S. Department of Energy, Office of Science, under contract DE-AC02-06CH11357. This work was also
supported by the National Science Foundation under Grants CCF-1619253, OAC-2003709, OAC-2034169, and OAC-2042084. We
acknowledge the Texas Advanced Computing Center for providing HPC resources that have contributed to the research
results reported within this paper.

\renewcommand*{\bibfont}{\small}
\printbibliography[]

\end{document}